\documentclass[12pt]{iopart}
\usepackage{iopams}
\usepackage{epsfig}
\begin{document}
\title{Critical Study on the Absorbing Phase Transition in a Four-State Predator-Prey Model in One Dimension}
\author{Su-Chan Park}
\address{Department of Physics, The Catholic University of Korea, Bucheon 420-743, Korea}
\eads{\mailto{spark0@catholic.ac.kr}}
\date{\today}
\begin{abstract}
In a recent letter [Chatterjee R, Mohanty P K, and Basu A,  2011 {\em J. Stat. Mech.} L05001], it was claimed that a four-state predator-prey (4SPP) model exhibits critical behavior distinct from that of directed percolation (DP). In this Letter, we show that the 4SPP in fact belongs to the DP universality class.
\end{abstract}
\maketitle
Absorbing phase transitions have been studied extensively as prototypical 
emergent phenomena of nonequilibrium systems and 
several universality classes have been
identified (for a review, see, e. g., Refs.~\cite{MD1999,H2000,O2004,L2004}). 
Among them, the directed percolation (DP) universality class
has been studied thoroughly both numerically and analytically. As a result,
a so-called `DP-conjecture'~\cite{J1981,G1982} has emerged since early
1980's, asserting that a model should belong to the DP class,
if the model exhibits a phase transition from an active, fluctuating
phase into a {\it unique} absorbing state, if the phase transition
should be characterized by a one-component order parameter,
if no quenched disorder is involved, if dynamic rules of the model 
are short-ranged processes,
and if the model has neither novel symmetry nor conservation. 
The validity of the conjecture has been supported by numerical
evidences. Although the theory suggests that the DP class is robust, 
an experimental realization of the DP class was successful only
recently~\cite{TKCS2007}. 

Theoretically, it has been questioned if each condition for the conjecture is 
necessary. Indeed, some of conditions are shown to be necessary.
It is by now firmly established that conditions of no quenched disorder and 
of short-range processes are necessary
for the DP critical behavior (for a review of non-DP critical behavior
by quenched disorder or by long-range interactions,
see, e. g., Ref.~\cite{O2004}). 
On the other hand, it turns out that certain models that violate some
of the above mentioned conditions still belong to the DP class.
At first, the condition for a {\it unique} absorbing state is
shown to be not necessary.  A prototypical model is
the pair contact process (PCP)~\cite{J1993}
which has infinitely many absorbing states (IMAS) but
belongs to the DP class.
Second, the condition of the absence of symmetry or conservation 
should be applied with care.
For instance, a parity conservation sometimes changes the
universality class as happens in 
branching annihilating random walks with even number of
offspring~\cite{TT1992,CT1996}, but sometimes it has no role as in
a model of diffusing particles with reaction dynamics,
$2A\rightarrow 0$ and $3A \rightarrow 5A$~\cite{KC2003,PP2009}.

More notably, the condition of a one-component order parameter is rather 
tricky because
this condition requires coarse-graining of microscopic models. 
In particular, the coarse-graining should result in 
reggeon field theory~\cite{CS1980}. A question in this direction arises if we 
have a systematic prescription to conclude that an absorbing phase transition
of a certain microscopic model is characterized by a one-component 
order parameter field theory. A naive answer would be 
the number of species in a microscopic model.
Here, by species is meant a group of particles whose dynamic
rules are same irrespective of position and time. 
However, there are counter-examples against this naive answer.
At first, certain models such as the Ziff-Gulari-Barshad 
model~\cite{ZGB1986} also exhibit the DP criticality,
although there are two different species in the model.  Second, to make
things even more complicated, certain single-species microscopic model may 
have multi-component order parameters in field theory.
To explain the meaning of the above sentence in a concrete way, let us 
consider single-species models with ``hybrid'' rules such as 
$n A \rightarrow 0$ and $ m A \rightarrow (m+1)A$ ($n,m\ge 1$) of diffusing 
$A$ particles~\cite{KC2003}, which show different critical behavior
depending on the values of $n$ and $m$.
If $m > n$, this model belongs to the DP class~\cite{KC2003}. However,
if $m=n$, the universality class to which this model belongs differs with
$n$. For the case of $m=n=1$, it still belongs to the DP class. For 
the case of $m=n=2$ which is also known as the
pair contact process with diffusion (PCPD) (for a 
review see, e. g., Refs.~\cite{HH2004,PP2008EPJB}) and for the 
case of $m=n=3$ which is also known as the
the triplet-contact process with diffusion (TCPD)~\cite{PHK2002},
these models do not share criticality with DP at least 
in dimensions higher than 1. 
Regarding the number of order parameter components, 
it was argued by studying the PCPD with biased 
diffusion~\cite{PHP2005a, PHP2005b} that the field theory of the PCPD should 
have at least two different order parameter components, which 
cannot be reduced to a local field theory with a single component 
in the asymptotic regime.
Failure of a single-component field theory for the PCPD~\cite{JvWDT2004}
also suggests the same conclusion arrived at in Ref.~\cite{PHP2005a}.
To summarize, the condition of single component order parameter
cannot be answered by simply looking at dynamic rules of a model. 

In this context, the conclusion in the recent Letter by 
Chatterjee et al.~\cite{CMB2011} is worth deep investigation.
In that Letter, it was asserted by numerical studies that
a four-state predator-prey (4SPP) model does not belong to the
DP class although the property of the model 
does not seem to deviate much from the conditions of 
the DP-conjecture. If the claim by Chatterjee et al.~\cite{CMB2011} were
right, a theory should be developed why
the 4SPP model does not belong to the DP class, which would
deepen our understanding of the DP class.
In this Letter, we will study the 4SPP model in detail.

Unfortunately, however, there is an apparent fallacy in Ref.~\cite{CMB2011}. 
The 4SPP model without diffusion was claimed to have only two absorbing states 
one of which is dynamically unstable. In fact, however, there are infinitely 
many absorbing states (see below). Because of this mistake, 
Chatterjee et al.~\cite{CMB2011} chose a wrong order
parameter, which eventually led them to locate a wrong critical point. 
In this Letter, we will show that the critical point in Ref.~\cite{CMB2011} 
is actually underestimated
and at the right critical point, the 4SPP model exhibits the DP scaling.

For completeness, we will begin with explaining the 4SPP model. 
In the 4SPP model, there are two
species; prey (called $A$) and predator (called $B$). At most
one individual from each species can reside at any lattice point, but two different
species can stay together at the same site. Hence, each site can have one of
four possible states; empty, $A$ alone occupied, $B$ alone occupied, and both 
$A$ and $B$ occupied. $A$ can spontaneously 
generate an offspring to its nearest neighbor, but $B$ cannot branch 
its offspring without consuming $A$. 
However, unlike $A$ species, predators also compete with each other for space and
if two predators are located in a row, both of them can be killed together and removed
from the system.
In this Letter, we restrict ourselves to a one-dimensional lattice system
with size $L$.  Periodic boundary conditions are always assumed.

\begin{figure}[t]
\centering
\includegraphics[width=0.7\textwidth]{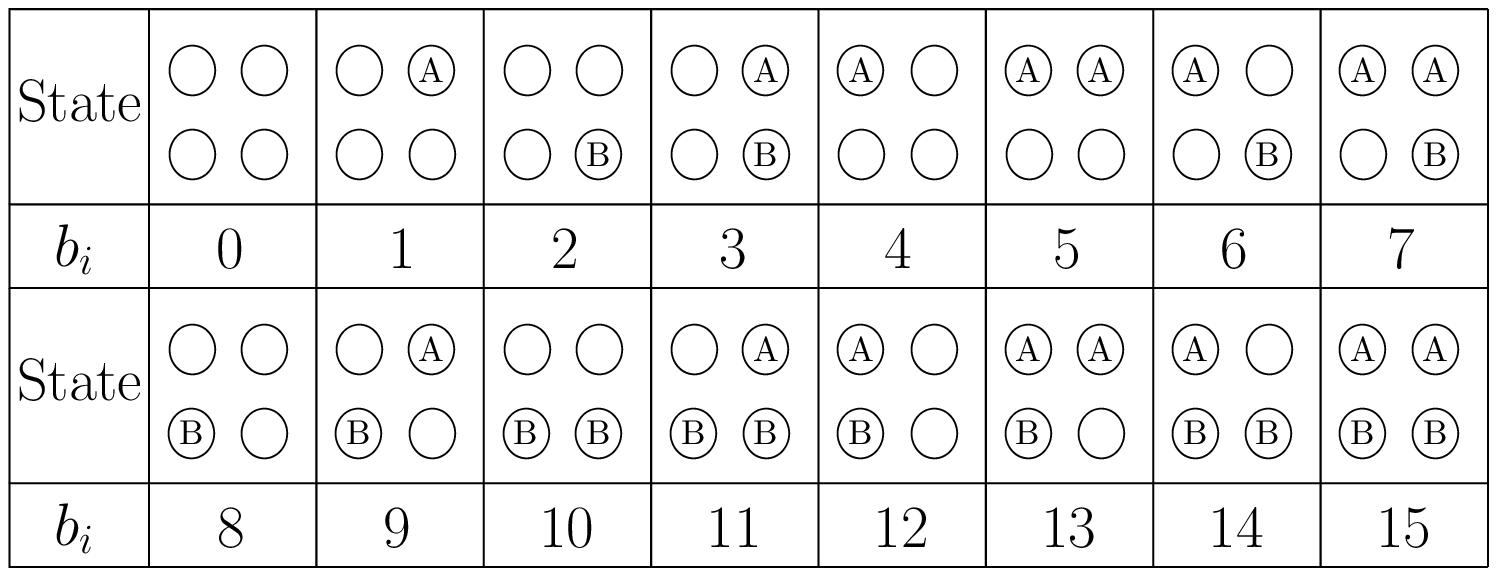}
\caption{\label{Fig:visual} The connection between the bond variable
$b_i$ and the (local) state at sites $i$ (circles in left columns) and $i+1$ (circles in right columns).
Circles in the top (below) rows indicate whether $A$ ($B$) species is present or not.
For example, the state corresponding to $b_i = 11$ is that at site $i$ only $B$ is
occupied and at site $i+1$, both $A$ and $B$ are occupied.}
\end{figure}

To describe the dynamic rules of the model in detail, we introduce a state 
variable $s_i$ at site $i$ ($1\le i \le L$) such that
$s_i = s_i^A + 2 s_i^B$, where $s_i^A$ ($s_i^B$) is the number of $A$ ($B$) species
at site $i$. Obviously, $s_i$ can take one of four possible values, 0, 1, 2, and 3.
We also assign a bond variable $b_i$ to a bond connecting $i$-th and $(i+1)$-th
sites such that $b_i = 4 s_i + s_{i+1}$ (recall the periodic boundary conditions).
The connection between a local configuration and a bond variable is depicted
in \Fref{Fig:visual}.

At first, we assume that both $A$ and $B$ are not diffusive. As in Ref.~\cite{CMB2011},
branching occurs in a biased way. In this case, the bond variables
changes by the following stochastic rules:
\begin{eqnarray}
\nonumber
4 \rightarrow 5 \textrm{ with rate } p,\quad&&
6 \rightarrow 7 \textrm{ with rate } p,\\
\nonumber
9 \rightarrow 10 \textrm{ with rate } r,\quad&&
10 \rightarrow 0 \textrm{ with rate } q,\\
\nonumber
11 \rightarrow 1 \textrm{ with rate } q,\quad&&
12 \rightarrow 13 \textrm{ with rate } p,\\
\nonumber
13 \rightarrow 14 \textrm{ with rate } r,\quad&&
14 \rightarrow \left \{ \begin{array}{l} 15 \textrm{ with rate } p,\\
4 \textrm{ with rate } q
\end{array}
\right .
\\
15 \rightarrow 5 \textrm{ with rate } q,
\label{Eq:rule}
\end{eqnarray}
where numbers on the left (right) hand side of the arrows signifies the bond 
variables before (after) a change with corresponding transition rates. 
$p$ is the transition rate of $A$ branching event,
$q$ is the transition rate of pair annihilation of $B$ species by competition,
and $r$ is the transition rate of $B$ branching event by consuming one $A$.
Since a configuration change can occur if a bond variable is an element
of the following set
\begin{equation}
$S=\{4, 6, 9, 10, 11, 12, 13, 14, 15\}$,
\label{Eq:active_set}
\end{equation}
we will refer to a bond with $b_i \in S$ as an active bond.

In simulations, we keep a list of active bonds at each time. 
We choose a time-increment $dt$ such a way that
\begin{equation}
0< dt \le \frac{1}{\textrm{max}(p+q,r)}.
\label{Eq:dt}
\end{equation}
Assume
that there are $N(t)$ active bonds at time $t$. Among $N(t)$ active bonds,
we choose one at random. If $b_i=4$, 6, or 12, $b_i$ increases by 1 with probability 
$p\,dt$ (branching of $A$ species). 
In case this change happens, $b_{i+1}$ also increases by 4.
If $b_i = 9$ or 13, $b_i$ increases by 1 with probability $r\,dt$ 
(branching of $B$ species by consuming one $A$). 
In case this change happens, $b_{i+1}$ also increases by 4.
If $b_i=10$, 11, or 15, $b_i$ decreases by 10 with probability $q\,dt$ (pair annihilation
of two predators). In case this change happens, $b_{i+1}$ ($b_{i-1}$) also decreases by 8 (2). If $b_i=14$, $b_i$ increases by 1 with probability 
$p\,dt$ (branching
of $A$ species) or decreases by 10 with probability $q\,dt$ 
(pair annihilation of two predators).
If one of these two events happens, the bond variables of its neighbors also 
changes accordingly. After an attempt of a configuration change, time increases by $dt/N(t)$
and the list of active bonds is to be updated. We terminate the above procedure
if time exceeds the predetermined maximum observation time or the system does
not have any active bonds. In the following, we choose the transition rates
in such a way that $dt$ can be chosen to be 1. In all simulations,
the initial condition is that $s_i = 3$ for all $i$.

Clearly, a configuration without any active bonds is absorbing in the sense that
no further configuration change can occur.  Hence, unlike the claim
in Ref.~\cite{CMB2011}, the 4SPP model without diffusion has infinitely many
absorbing states (any configuration with only isolated predators
is absorbing). Because Chatterjee et al.~\cite{CMB2011}
ignored the absorbing nature of isolated predators, the density of 
predators is wrongfully chosen as an order parameter.
Since the 4SPP is an IMAS model and the density of predators
need not be zero in the absorbing state, the critical point reported in 
Ref.~\cite{CMB2011} is not the correct one. 
Actually, the density of active bonds defined as 
\begin{equation}
\rho(t) \equiv \frac{N(t)}{L},
\label{Eq:order}
\end{equation}
where $N(t)$ is the number of active bonds (see above) at time $t$ and 
$L$ is the system size,
can characterize correctly the onset of phase transition.

\begin{figure}[t]
\centering
\includegraphics[width=\textwidth]{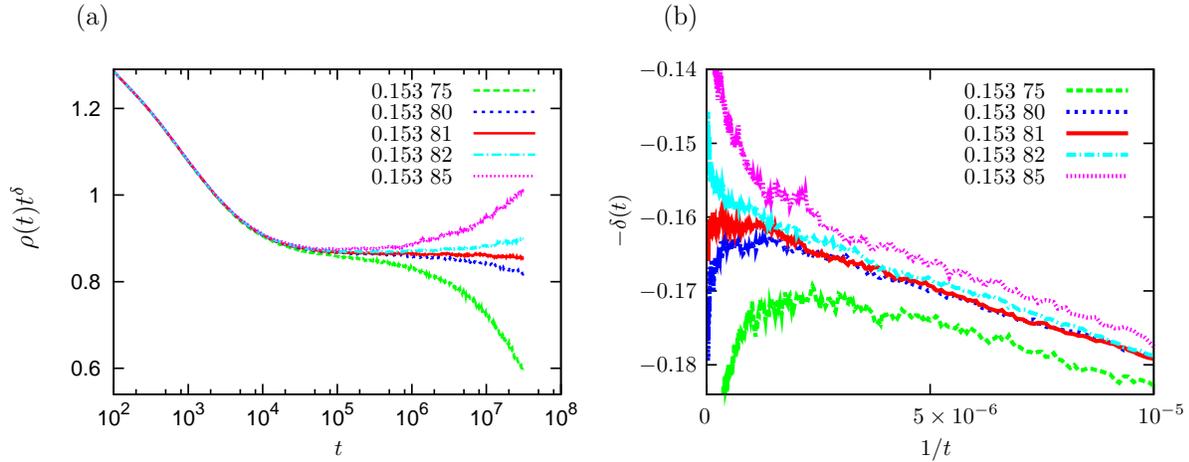}
\caption{\label{Fig:inf} (a) Semi-logarithmic plots of $\rho(t) t^\delta$ vs $t$ with DP critical exponent
$\delta = 0.1594$ for $p=0.153~75$, 0.153~80, 0.153~81, 0.153~82, and 0.153~85 with
$q=0.02$ and $r=0.9$. (b) Plots of effective exponents $-\delta(t)$ as a function
of $1/t$ for the simulation results in (a). 
Since the effective exponent veers down (up)
in the absorbing (active) phase, the critical point is estimated as $p_c = 0.153~81(1)$.
Moreover, the effective exponent approaches to the DP critical exponent 0.1594 with
linear behavior in $1/t$.
}
\end{figure}
At first, we study how $\rho(t) t^\delta$ with $\delta \simeq 0.1594$ 
(DP exponent~\cite{J1999}) behaves near criticality. To compare our results with those in Ref.~\cite{CMB2011},
we fix $r=0.9$ and $q=0.02$ and the criticality is 
found by varying $p$.  We simulated the system with size $L=2^{20}$ 
and the number of independent runs, 
for instance, for $p=0.153~81$ is 400. Up
to the observation time $t_\mathrm{max} \approx 3 \times 10^7$, all
simulated systems did not fall into one of absorbing states, which minimally
guarantees that the finite size effect is not operative.
Note that the system size and the observation time in this Letter are much
larger than those in Ref.~\cite{CMB2011}.
In \Fref{Fig:inf}(a), semi-logarithmic plots of
the density behavior for
various $p$'s are depicted. From $t=10^5$, one can see that the curve for
$p=0.153~81$ becomes almost constant, and the curve for $p=0.153~80$ (0.153~82)
veers down (up), which indicates that the critical point is 
$p_c = 0.153~81(1)$ with the number in parentheses to be the error of the last digit. 

To be more systematic, we also study the effective exponent defined as ($b$ is
a certain fixed constant)
\begin{equation}
-\delta(t) \equiv \frac{\ln \rho(t)/\rho(t/b)}{\ln b}.
\end{equation}
In the active phase the effective exponent should veer up because
$\rho(t)$ should saturate to a finite number. 
In the absorbing phase, on the other hand, the effective
density should veer down because $\rho(t)$ decays exponentially 
(or at least faster than the critical decay). 
And only at criticality, $\delta(t)$ will approach to
the critical exponent $\delta$ as $t\rightarrow \infty$.
Note that if the leading corrections to scaling takes the $1/t$ form, that is,
if $\rho(t) \approx t^{-\delta} ( 1 + a/t)$ in the asymptotic regime 
with a constant $a$, which is
the case in many models belonging to the DP class, $\delta(t)$ at criticality
approaches to $\delta$ linearly if $\delta(t)$ is drawn as a function of
$1/t$.

We analyze the effective exponent using the same data in \Fref{Fig:inf}(a)
with $b=10$.  The resulting plots of the effective exponents are shown 
in \Fref{Fig:inf}(b).  Again we conclude that 
the critical point is $0.153~81(1)$ and the effective exponent approaches to
the DP critical exponent 0.1594 linearly in $1/t$.

\begin{figure}[t]
\centering
\includegraphics[width=\textwidth]{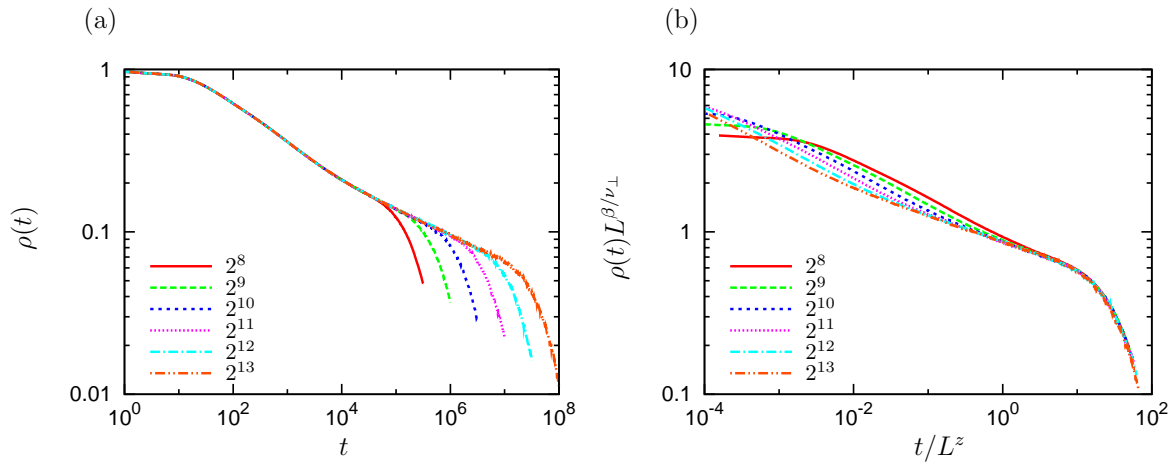}
\caption{\label{Fig:fin} (a) Plots of $\rho_s(t)$ vs $t$ at criticality ($q=0.02$,
$r=0.9$, $p=0.153~81$)
for $L=2^8$, $2^9$, $2^{10}$, $2^{11}$, $2^{12}$, and $2^{13}$. (b) Plots
of $\rho_s(t)L^{\beta/\nu_\perp}$ vs $t/L^z$ at criticality with DP exponents
$\beta/\nu_\perp \approx 0.252$ and $z\approx 1.58$. As anticipated from \Fref{Fig:inf}(a),
the data collapse becomes good when the time that the finite size effect appears
is larger than $10^5$.
}
\end{figure}
To make our point stronger, we also study the finite size scaling. To this end,
we simulated the system at criticality $p=0.153~81$ with varying system sizes
from $L=2^8$ to $L=2^{13}$. \Fref{Fig:fin}(a) shows the density decay behavior
for different system size.  Since the order parameter takes the form in the scaling regime
\begin{equation}
\rho(t) = F(t/L^z) L^{-\beta/\nu_\perp},
\end{equation}
where $z$, $\nu_\perp$, and $\beta$ are critical exponents and $F$ is a scaling function,
plots of $\rho(t) L^{\beta/\nu_\perp}$ vs $t/L^z$ for different $L$ should collapse
into a single curve in the scaling regime. \Fref{Fig:fin}(b) shows a nice scaling collapse, 
using $z = 1.58$ and $\beta/\nu_\perp = 0.252$ (DP exponents~\cite{J1999}), which again confirms that
the 4SPP belongs to the DP class. 

Up to now, we assume that both $A$ and $B$ species are not diffusive and we show 
that the 4SPP model with non-diffusive species does belong to the DP class.
Now we will study if diffusive motion can affect the universal behavior.
In Ref.~\cite{CMB2011}, though no explicit numerical results are shown, the
4SPP model with diffusion is also claimed not to belong to the DP class. 
From the experience of the PCPD and the TCPD,
we know that diffusion introduced to the model with IMAS may change
the universality class. Quite interestingly, the exponents reported
in Ref.~\cite{CMB2011} look very similar to those in Ref.~\cite{NP2004} 
(in particular
see the reported exponents for $r=0.5$ in Table II of Ref.~\cite{NP2004}), which
studies a variant of the PCPD. 
Hence we investigated if Chatterjee et al. indeed observed the PCPD behavior 
from the 4SPP model with diffusion.
Since the one-dimensional PCPD is known to have strong corrections to
scaling and the true scaling behavior has not yet been properly observed, 
Ref.~\cite{CMB2011} might have observed the intermediated behavior of the PCPD,
if the 4SPP model with diffusion is PCPD-like.
To check this possibility, we study the role of biased diffusion,
following the suggestion in Ref.~\cite{PHP2005a}.
That is, if the 4SPP model with biased diffusion
shows the same critical behavior as the driven PCPD~\cite{PHP2005a}, 
one can conclude that
the 4SPP with unbiased diffusion will share criticality with the PCPD.

Let $D_A$ ($D_B$) be a transition rate with which $A$ ($B$) moves to its right nearest 
neighbor. As reasoned above, movement of both species are assumed to be biased.
For completeness, we fully write the transition rate including diffusion:
\begin{eqnarray}
\nonumber
4\rightarrow  \left \{
\begin{array}{l}
5 \textrm{ with rate } p,\\
2 \textrm{ with rate } D_A,
\end{array} \right .
\quad&&
6\rightarrow  \left \{
\begin{array}{l}
7 \textrm{ with rate } p,\\
3 \textrm{ with rate } D_A,
\end{array} \right .
\\
\nonumber
8\rightarrow 2 \textrm{ with rate } D_B,\quad&&
9 \rightarrow 
\left \{
\begin{array}{l}
10 \textrm{ with rate } r,\\
3 \textrm{ with rate } D_B,
\end{array} \right .
\nonumber\\
10 \rightarrow 0 \textrm{ with rate } q,\quad&&
11 \rightarrow 1 \textrm{ with rate } q,
\nonumber
\\
12 \rightarrow 
\left \{
\begin{array}{l}
13 \textrm{ with rate } p,\\
9 \textrm{ with rate } D_A,\\
6 \textrm{ with rate } D_B,
\end{array}\right .
\quad&&
13 \rightarrow 
\left \{ \begin{array}{l}
14 \textrm{ with rate } r,\\
7 \textrm{ with rate } D_B
\end{array}
\right .
\nonumber 
\\
14 \rightarrow \left \{ \begin{array}{l} 15 \textrm{ with rate } p,\\
4 \textrm{ with rate } q,\\
11 \textrm{ with rate } D_A,
\end{array}
\right .
\quad&&
15 \rightarrow 5 \textrm{ with rate } q.
\label{Eq:rule_D}
\end{eqnarray}
Now time-increment $dt$ is chosen to be in the range
\begin{equation}
0< dt \le \frac{1}{\mathrm{max} 
\left (
p+D_A+D_B, p+q+D_A, r+D_B
\right )
}.
\end{equation}
In practice, we chose transition rates such that $dt=1$ can be used
near criticality.

If $D_A \neq 0$ and $D_B = 0$, the change of the universality class is not expected, because
the model still has IMAS and no other symmetry and/or
conservation emerges. The order parameter defined in \Eref{Eq:order} is
again used for $D_B =0$ because
the classification of active bonds does not change. Indeed, as we will see 
later, the diffusion of $A$ does not affect
the universal behavior.

The structure of the absorbing states is significantly affected
by diffusion of $B$. If $D_B\neq 0$, an isolated predator can diffuse around until it meets
another $A$ or $B$. Hence the number of absorbing states shrinks to 3; 
the completely empty configuration, 
a configuration with a single predator without any prey, and 
the $A$ fully occupied configuration without $B$. 

At first sight, the diffusion of 
predators seems to have a similar role of the diffusion in the PCPD. That is, an isolated predator
neither dies out nor branches another offspring until it meets another predator or prey.
On the other hand, death and branching events are mediated by different local 
states, that is, death requires two predators but branching requires one predator
and one prey, although in the PCPD both branching and annihilation occur by the
same local configuration (that is, by a pair).
Actually, whether branching and annihilation is mediated
by the same local configuration or not seems to be an important factor. 
Unlike the PCPD and the TCPD, the universal behavior of models
with hybrid rules with $m > n$
remains intact despite of diffusion~\cite{KC2003,PP2006,PP2007}.
In this regard, the 4SPP with diffusing predators 
is expected to belong to the DP class irrespective of whether the diffusion
is biased or not, because this model is similar 
to the model with hybrid-rules of $m>n$ rather than the PCPD.
We will show that the 4SPP model indeed belongs to the DP class.
\begin{figure}[t]
\centering
\includegraphics[width=\textwidth]{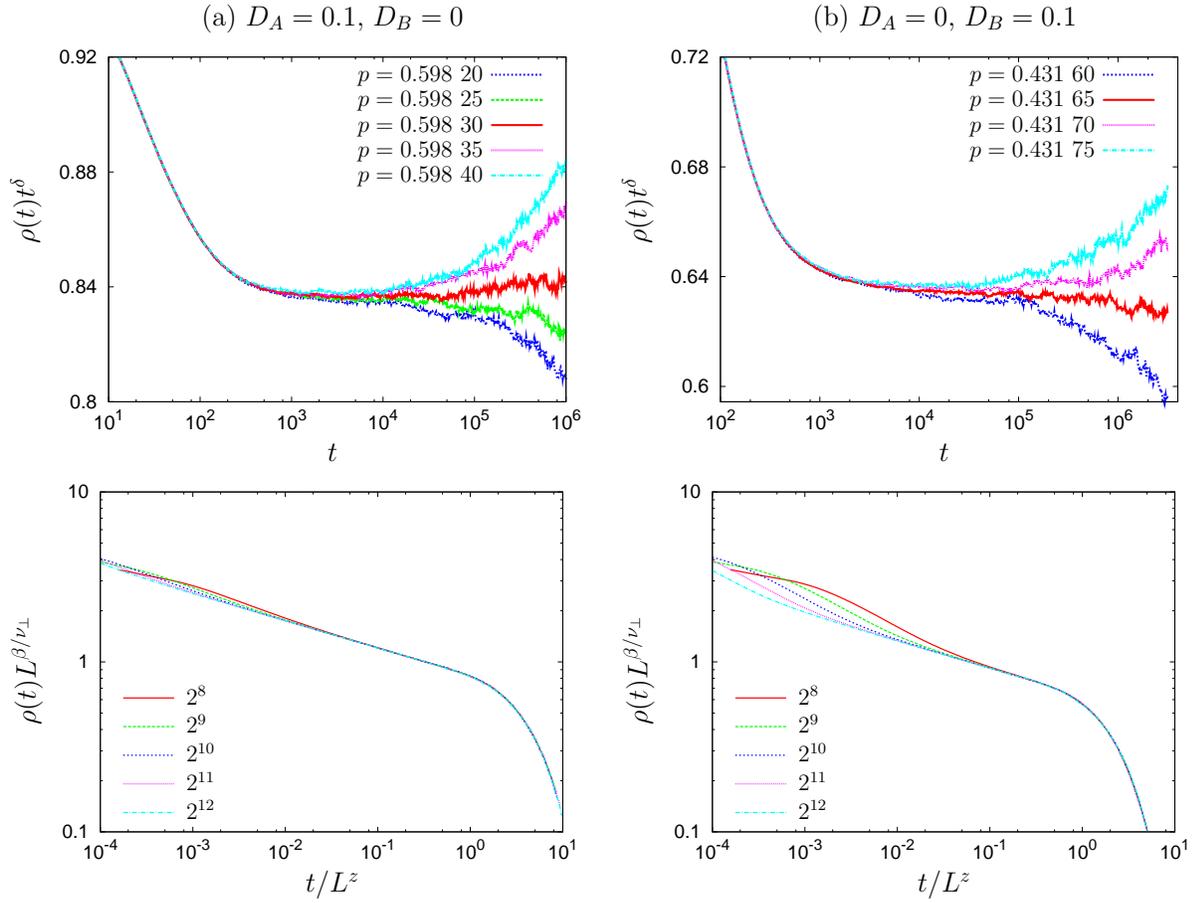}
\caption{\label{Fig:mov} Numerical Study of critical phenomena 
 for (a) $D_A = 0.1$ and $D_B = 0$ (two figures in left column) and 
for (b) $D_A = 0$ and $D_B = 0.1$ (two figures in right column).
$q = 0.1$ and $r=0.9$ is used. $\delta = 0.1594$, $z=1.58$, and $\beta/\nu_\perp
=0.252$ (DP exponents) are used. From the two figures in above row, we conclude
that $p_c = 0.592~30(5)$ for (a) and $p_c = 0.431~65(5)$ for (b).
Two figures in lower row shows the finite size scaling collapse at corresponding criticality, which concludes that both models belong to the DP class.
}
\end{figure}

Since the diffusion of $A$ is not expected to affect the critical behavior, we study
only two cases; one is for $D_A = 0.1$ and $D_B = 0$ and the other is for
$D_A = 0$ and $D_B = 0.1$. $q$ and $r$ are fixed to be $0.1$ and $0.9$, respectively,
and $p$ is chosen to be the tuning parameter as before.
As in the study of the non-diffusive 4SPP model, we study both the critical 
decay
and the finite size scaling. The resulting analyses are summarized in \Fref{Fig:mov}.
For the model with $D_A=0.1$ and $D_B = 0$ which has IMAS, the critical
point is found to be $p_c = 0.598~30(5)$ by studying 
$\rho(t) t^\delta$ with the DP exponent $\delta \simeq 0.1594$. As expected,
the critical density decay shows  DP behavior. The finite
size scaling collapse with the DP exponents again confirms that
the 4SPP with $D_A = 0.1$ and $D_B=0$ does belong to the DP class 
[see \Fref{Fig:mov}(a)].

For the case of $D_A = 0$ and $D_B = 0.1$, 
a bond is active if a bond variable is an element
of the following set
\begin{equation}
S' = S \cup \{8\} = \{4, 6, 8, 9, 10, 11, 12, 13, 14,15\},
\end{equation}
and now $N(t)$ is the number of these active bonds and the order parameter
$\rho(t)$ is the density of such active bonds.
In this case, the predator density also can be an order parameter.
We again analyze the critical decay of the order parameter and 
the finite size scaling.
At first, 
the density decay at $p_c = 0.431~65(5)$ is also well described by the 
DP exponent. 
The finite size scaling confirms that the 4SPP belongs
to the DP class [see \Fref{Fig:mov}(b)]. 
Hence, we conclude that irrespective of diffusion, the 4SPP does belong to the DP class.

To summarize, we studied the four-state predator-prey model introduced
in Ref.~\cite{CMB2011} with and without diffusion. Unlike the conclusion made
in Ref.~\cite{CMB2011}, we found that the 4SPP does belong to the DP class regardless
of whether prey and/or predators are diffusive.

The author thanks Jae Dong Noh for calling his attention to Ref.~\cite{CMB2011}.
He also acknowledge the hospitality of
Institut f\"ur Theoretische Physik, Universit\"at zu K\"oln, Germany and 
support under DFG grant 
within SFB 680 \textit{Molecular Basis of Evolutionary Innovations}
during the final stages of this work.
This work was supported by the Basic Science Research Program
through the National Research Foundation of Korea
(NRF) funded by the Ministry of Education, Science
and Technology (Grant No. 2011-0014680).
The computation of this work was supported by 
Universit\"at zu K\"oln, Germany.

\end{document}